\documentclass[fleqn,preprintnumbers,amsmath,showkeys,amssymb,prb,12pt]{revtex4}
\usepackage{graphicx}
\usepackage{dcolumn}
\usepackage{bm}
\usepackage{epsf}
\newcommand{\newc}{\newcommand}
\newc{\beq}{\begin{equation}}
\newc{\eeq}{\end{equation}}
\newc{\beqa}{\begin{eqnarray*}}
\newc{\eeqar}{\end{eqnarray}}
\newc{\beqar}{\begin{eqnarray}}
\newc{\eeqa}{\end{eqnarray*}}
\newc{\bd}{\begin{displaymath}}
\newc{\ed}{\end{displaymath}}
\newc{\mbf}{\mathbf}
\begin{document}
\title{Gauge momentum operators for the Calogero-Sutherland model \\
with anti-periodic boundary condition\footnote[0]{12pt font, double-spacing used for easy legibility.}}
\author{Arindam Chakraborty}
\affiliation{Department of Physics, Jadavpur University, Calcutta 700 032, India}
\author{Subhankar Ray}\email{sray@phys.jdvu.ac.in}
\affiliation{Department of Physics, Jadavpur University, Calcutta 700 032, India}
\affiliation{C.N. Yang Inst. for Theoretical Physics, Stony Brook, NY 11794, USA}
\author{J. Shamanna}
\email{jlsphy@caluniv.ac.in}
\affiliation{Physics Department, University of Calcutta,
Calcutta 700029, India}

\begin{abstract}
The integrability of a classical Calogero systems with anti-periodic
boundary condition is studied. This system is equivalent to the
periodic model in the presence of a magnetic field. Gauge momentum 
operators for the anti-periodic Calogero system are constructed.
These operators are hermitian and simultaneously diagonalizable with 
the Hamiltonian. A general scheme for constructing such momentum operators for
trigonometric and hyperbolic Calogero-Sutherland model is proposed.
The scheme is applicable for both
periodic and anti-periodic boundary conditions. 
The existence of these momentum
operators ensures the integrability of the system.
The interaction parameter $\lambda$ is
restricted to a certain subset of real numbers.
This restriction is in fact essential for the construction of the
hermitian gauge momentum operators. 
\end{abstract}
\maketitle

\setcounter{equation}{0}
\setcounter{page}{1}

\section{Introduction}
The study of integrable and solvable quantum
many-body problem has become an active field of research with many 
applications in various branches of physics and mathematics. 
One dimensional models with two-body inverse square long-range
interaction are of great physical interest as they are extensively 
used to estimate the physical feature of several condensed 
matter systems, e.g., quantum 
hall effect \cite{hald81}, Luttinger liquid \cite{hald81b}. They also
have deep connections with Yang-Mills 
theories \cite{gor94, mina94},
soliton theory \cite{poly95}, random matrix models\cite{dyson62},
multivariable orthogonal polynomials \cite{jack69}, quantum gravity and 
black holes \cite{gibbons99}. In systems of physical interest,
these one-dimensional models with appropriate inverse square long-range 
interaction are exactly solvable due to the highly restrictive 
spatial degrees of freedom. 
The spatial restriction however, introduces large quantum fluctuations
resulting in failure of the mean field approach which works well in higher
dimensional systems. 

Several integrable models in one dimension
have been proposed and constructed using the inverse square potential.
The Calogero-Sutherland model (CSM)\cite{cal62,suth71} is one such model 
with applications in several physical systems.
This model provides a clear  explanation of fractional
exchange and exclusion statistics. 
In most other one-dimensional models, the definition of fractional exchange 
statistics is rather obscure and incomplete.
For spinless CSM, the fractional exchange statistics can be formulated
in the language of first quantization by using the one 
dimensional analogue of Chern-Symon Gauge theories \cite{poly92}. 

The CSM is also useful in the study of the fractional exclusion
statistics based on the generalization of the exclusion principle 
\cite{hald91}. The exclusion statistics is usually interpreted in terms of real, 
pseudo and quasi momenta which describe the particle and hole type 
excitations of one dimensional systems \cite{ha94}. 
In Calogero systems it is observed that two neighboring
pseudo-momenta are always separated by a number that depends on a
statistical parameter present in the Hamiltonian.
The study of such particle and hole type excitations is important 
for constructing the basic thermodynamic functions of a system. 
In addition, the CSM Hamiltonian can be used to construct an effective 
low energy model for anyons following Luttinger 
liquid theory \cite{hald81b}. For integer values of interaction parameter,
this type of one dimensional anyon system is equivalent to a coupled system
of left and right moving edge-states of fractional quantum Hall effect
\cite{poly89}.

The $N$-particle Hamiltonian of a general Calogero model represents 
a system of spinless non-relativistic particles interacting through a 
two-body potential and may be written as,
\beq\label{hamilton}
H_N =\sum_{j=1}^N{\partial_j}^2-\lambda(\lambda-1)
\sum_{\substack{j,k \\j\neq k}}U(x_{jk})
=\sum_{j=1}^N{\partial_j}^2-\lambda(\lambda-1)
{\sum_{j,k}}^\prime U(x_{jk})
\eeq
where, prime over the summation sign implies that the terms with $j=k$ 
are omitted. 

The two-body long-range potential is represented by $U(x_{jk})$,
where $x_{jk} = x_j - x_k$ is the distance between particles at $j$-th
and $k$-th sites
and $\lambda$ is a dimensionless interaction parameter. 
The two-body potential $U(x)$ is an even function and under periodic boundary 
condition has a general expression in terms of Weierstrass 
elliptic function \cite{wat}.
This may be further reduced to the trigonometric, hyperbolic 
and rational function
by means of a limiting procedure \cite{ols83}. The original version of the
Calogero model assumed a two-body inverse square potential. The model was
shown to be integrable by Calogero and Perelomov \cite{cal75,perel77} using
quantum Lax formulation and its explicit integration was
performed by Krichever \cite{krich80}.
The existence of a complete set of mutually commuting momentum
operators that commute with the Hamiltonian as well, also establishes
the integrability of this model. The CSM with periodic boundary condition
has been well studied in literature 
\cite{jack69,suth71,bern93,cad01,bela84}.
During the past decades the CSMs (both classical and spin system) have been
actively explored in a variety of ways including the exchange operator
formalism (EOF)\cite{mina93}, the Dunkl operator approach\cite{bern93},
reduction by discrete symmetries \cite{poly99} and construction of Lax-pair
\cite{hika93}. The operators used in EOF or
quantum Lax formulation are associated with certain types of root systems.
The Calogero type models may be obtained from the projection of free
motion on a higher dimensional manifold.

This article investigates the Calogero 
systems for trigonometric
and hyperbolic types of long-range interactions with anti-periodic 
boundary conditions \cite{habook}. Such systems are equivalent to
a periodic model in the presence of an external magnetic field.  
The trigonometric form of the Hamiltonian is obtained by mapping 
the one dimensional chain of particles on a circular ring. The hyperbolic 
version of the model is derived by expressing the
potential in terms of Weierstrass $\wp$-function \cite{cop}. The integrability
is established by constructing gauge momentum operators that are hermitian
and are simultaneously diagonalizable with the Hamiltonian.
A general scheme for constructing such momentum operators is proposed, for 
trigonometric and hyperbolic types of Calogero systems with
both periodic and anti-periodic boundary
conditions. The existence of these commuting momentum 
operators ensures the integrability of the models under consideration.
It is also shown in this article, that for periodic boundary
condition the gauge momentum
operator exists for any real values of $\lambda$ whereas for anti-periodic 
case there is a restriction on the range of allowed values of $\lambda$.
Moreover, for bosonic ($\lambda=0$)
and fermionic ($\lambda=1$) limits, both periodic and anti-periodic 
models have identical spectra but 
this similarity is absent for any other value of $\lambda$. 
\section{The Calogero-Sutherland model with trigonometric type interaction}
\subsection{Trigonometric CSM with periodic boundary condition}
Let us first consider the general CSM; a one dimensional
chain of classical particles  with inverse square long-range interaction.
The topological representation of this one dimensional 
chain is simply a circular ring. 
In the absence of a magnetic field, a particle transported adiabatically
around the ring an integral number of times, does not take up any phase 
factor, and hence the eigenfunctions retain their initial form. 
Thus, the pairwise interaction summed over all possible pairs, around a circle
of circumference $L$, an infinite number of times 
($\nu\rightarrow \infty$) is given as,
\beq\label{pair_int}
\lim_{\nu\rightarrow \infty}\sum_{n=-\nu}^{\nu} \frac{1}{(x+nL)^2}=
\frac{1}{[d(x)]^2} ,
\eeq
where $x$ is the distance along the arc of the circle, between 
the particles at the $j$-th and $k$-th sites, and $d(x_{jk})$ is the
the chord distance between them. 
This chord length is given by, (See Fig. 1)
\beq\label{chord}
d(x) = \frac{L}{\pi} \sin \left( \frac{\pi x}{L} \right).
\eeq
Hence,
\bd
U(x) = \frac{\pi^2}{L^2}\frac{1}{\sin^2\left(\frac{\pi x}{L}\right)}
\ed
\begin{figure}[h]
\resizebox{!}{2.0in}
{\hskip 1cm \includegraphics{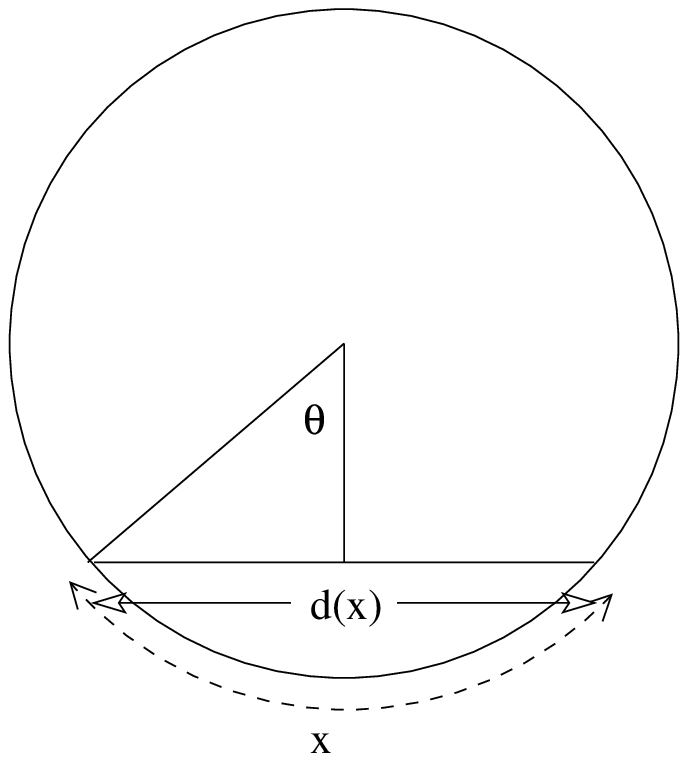}}
\resizebox{!}{2.0in}
{\hskip 1cm \includegraphics{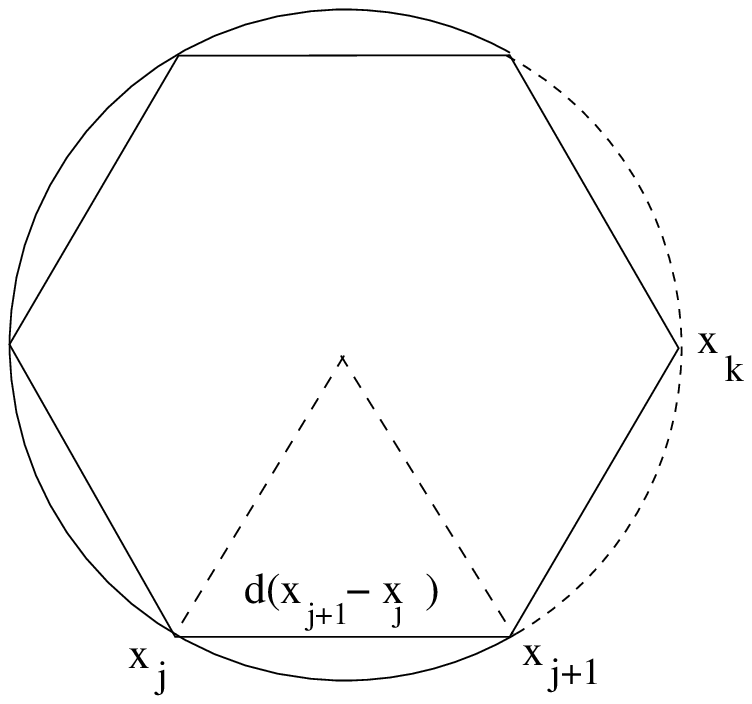}}
\caption{Interparticle distances $d(x)$ and $d(x_{j+1}-x_j)$ for particles on a circular chain}
\end{figure}
The Hamiltonian then is given by, 
\beq\label{htrig1}
H_N=\sum_{j=1}^{N}{\partial_j}^2-\lambda (\lambda-1)
\frac{\pi^2}{L^2} {\sum_{j,k}}^\prime \frac{1}{\sin^2{(\frac{\pi}{L}x_{jk})}} .
\eeq
Using standard trigonometric identity, Eq.(\ref{htrig1}) becomes,
\beq\label{htrig12}
H_N = \sum_{j=1}^{N}{\partial_j}^2-\frac{\pi^2}{4L^2}\lambda(
\lambda-1) {\sum_{j,k}}^\prime \left(\frac{1}{\sin^2{(\frac{\pi}{2L}x_{jk})}}+
\frac{1}{\cos^2{(\frac{\pi}{2L}x_{jk})}}\right) .
\eeq
Making a change of variable $(\pi/2L) x_j \rightarrow x_j$
and rescaling the Hamiltonian as $(4L^2/\pi^2)H \rightarrow H$, we get 
the trigonometric Hamiltonian with periodic boundary condition as follows
\beq\label{htrig2}
H^{t^+}_N=\sum _{j=1}^{N}{\partial_j}^2-\lambda
(\lambda-1){\sum_{j,k}}^\prime\left(\frac{1}{\sin^2{(x_{jk})}}+
\frac{1}{\cos^2{(x_{jk})}}\right) .
\eeq
\subsection{Trigonometric CSM with anti-periodic boundary condition}\label{tacsm}
The two-body interaction term in CSM Hamiltonian takes a different form when
the boundary condition is changed to anti-periodic. The change in 
boundary condition signifies a change of certain symmetry consideration 
in the underlying algebraic structure of the model. 
A general twisted boundary condition arises when a magnetic field is applied
transverse to the one dimensional ring considered above.
Then, if a particle is transported adiabatically around the entire 
system $n$ number of times, it picks up a net phase $\exp{(in\phi)}$. \\
The total pairwise interaction summed over all possible pairs
an infinite number of times around a circle of circumference $L$ is given by,
\beq\label{antipair_int}
\sum_{n=-\infty}^{+ \infty} \frac{\exp(i \phi n)}{(x+nL)^2} .
\eeq
The above summation can be evaluated if we make the choice,
\beqar\label{phi}
&& \phi=2\pi p/q, \hskip 0.3cm \mbox{with} \hskip 0.3cm p, \,q 
\hskip 0.3cm \mbox{relative primes; and }
\nonumber \\
&& n=jq+k, \hskip 0.3cm \mbox{with} \hskip 0.3cm j, \, k 
\hskip 0.3cm \mbox{integers such that} \hskip 0.3cm 
-\infty < j < +\infty, \hskip 0.3cm 
\mbox{and} \hskip 0.3cm 0 \le k \le (q-1).
\eeqar
Using the above, the interaction term becomes,
\beq\label{antisum}
\sum _{n=- \infty}^{+\infty}\frac{\exp{(i\phi n)}}{(x+nL)^2} =
\sum _{k=0}^{q-1} \sum _{j=- \infty}^{+\infty}
\frac{\exp(i 2 \pi pj) \exp(i 2 \pi p k/q)}{\left[(x+kL)+(qL)j \right] ^2}
=\sum _{k=0}^{q-1}\frac{\exp{(i 2\pi pk/q)}}
{\left[(qL/\pi) \sin[(\pi(x+kL))/qL]\right]^2}
\eeq

The last expression represents an interaction when the system is 
subjected to a general twisted boundary condition. The model can be viewed as
a system of interacting particles residing on a circle with circumference
$qL$. For $p/q=1/2$ this
corresponds to an anti-periodic boundary condition \cite{habook}.
In this case the sum in Eq.(\ref{antisum}) may be written as
\beqar\label{antiu}
\sum _{k=0}^{1}\frac{\exp{(i \pi k)}}
{\left[(2L/\pi) \sin[(\pi(x+kL))/2L]\right]^2}
&=&\frac{1}{\left[(2L/\pi) \sin[(\pi x)/2L]\right]^2}-
\frac{1}{\left[(2L/\pi) \sin[(\pi(x+L))/2L]\right]^2} \nonumber \\
&=&\frac{\pi^2}{4 L^2 \sin^2[(\pi x)/2L]}-
\frac{\pi^2}{4 L^2 \cos^2[(\pi x)/2L]} .
\eeqar
The above expression represents the potential term for the Calogero system
with anti-periodic boundary condition.
Making a change of variable $(\pi/2L) x_j \rightarrow x_j$ and 
rescaling the Hamiltonian as $(4L^2/\pi^2)H \rightarrow H$, 
the trigonometric Hamiltonian with anti-periodic boundary 
condition, denoted by $H_N^{t^-}$, is written as
\beq\label{htrig3}
H_N^{t^-}=\sum _{j=1}^{N}{\partial_j}^2-\lambda
(\lambda-1){\sum_{j,k}}^\prime\left(\frac{1}{\sin^2{(x_{jk})}}-
\frac{1}{\cos^2{(x_{jk})}}\right) .
\eeq
 
The two Hamiltonians in Eq.(\ref{htrig2}) and Eq.(\ref{htrig3})
may be expressed in a compact form as,
\beq\label{htrig_gen}
H_N^{t^{\pm}}=\sum _{j=1}^{N}{\partial_j}^2-\lambda
(\lambda-1){\sum_{j,k}}^\prime[V(x_{jk})\pm R(x_{jk})] ,
\eeq
where, the upper sign is for the periodic and the lower sign is for the 
anti-periodic case. $V(x)$ and $R(x)$ are even functions of $x$.

\subsection{Construction of commuting gauge momentum operators} 
In order to construct the gauge momentum operators we introduce the exchange 
operator $\Lambda_{jk}$ which preserves the function space under exchange 
of coordinates of the particles. The exchange operator $\Lambda_{jk}$ 
has the following properties, 
\begin{enumerate}
\item[1.]$\Lambda_{jk}f(x_1,..,x_j,..,x_k,.., x_N)=
f(x_1,..,x_k,..,x_j,.., x_N)$ .
\item[2.]$\Lambda_{jk}=\Lambda_{kj}$ .
\item[3.]$\Lambda_{jk}^2=1$ .
\item[4.]$\Lambda_{ij}\Lambda_{jk}=\Lambda_{ik}\Lambda_{ij}
=\Lambda_{jk}\Lambda_{ik}$ .
\item[5.]$\Lambda_{ij}\Lambda_{kl}=\Lambda_{kl}\Lambda_{ij}$ .
\item[6.]$\Lambda_{jk} x_k = x_j$ .
\end{enumerate}
Let us define 
\beq\label{vr}
v(x)=\frac{1}{2}\frac{d}{dx}\ln V(x),
\hskip 2cm
r(x)=\frac{1}{2}\frac{d}{dx}\ln R(x).
\eeq
It can be easily shown that $r(x) \cdot v(x)=-1$ . 
From the definition it is clear that $v(x)$ and $r(x)$ are odd functions of 
$x$.
Using the exchange operator 
$\Lambda_{ij}$, we define the gauge momentum operators
$\{d_j\vert j=1\dots N\}$ in terms of the functions $v(x)$ and $r(x)$  as
\beq\label{moment_def}
d_j={\partial _j}+\mu_1(\lambda)\sum_{\substack{k\\ k\neq j}}
v(x_{jk})\Lambda_{jk}
+\mu_2(\lambda)\sum_{\substack{k\\ k\neq j}}r(x_{jk})\Lambda_{jk} 
\eeq    
$\mu_1(\lambda)$and $\mu_2(\lambda)$ being real functions of $\lambda$. 
To ensure integrability, we require the
momentum operators to satisfy the following relations,
\beq\label{hc0}
[\Lambda_{ij},d_k]=0  .
\eeq
\beq\label{hc1}
\sum_{j=1}^N d_j^2= H_N^{t^{\pm}} \, + \, \mbox{constant}  .
\eeq
\beq\label{hc2}
[d_j,d_k]=0 .
\eeq
\beq\label{hc3}
[d_j,H_N^{t^{\pm}}]=0 .
\eeq
This implies that $r(x)$ and $v(x)$ must obey the following restrictions 
\beqar\label{const}
\frac{d}{dx}v(x)=V(x) \; , \hskip 2cm v^2(x)=V(x)+ \mbox{constant} \\
\frac{d}{dx}r(x)=R(x) \; , \hskip 2cm  r^2(x)=R(x)+ \mbox{constant}
\eeqar
Making use of the properties of the momentum operators (Eq.(\ref{hc0})
-Eq.(\ref{hc3})) and Eq.(\ref{const}), the momentum operators 
$\{d_j\vert j=1\dots N\}$ in Eq.(\ref{moment_def}) can be written as,
\beq\label{moment}
d_j={\partial _j} - \mu_1(\lambda)\sum_{\substack{k\\k\neq j}}
\cot(x_{jk})\Lambda_{jk}
+\mu_2(\lambda)\sum_{\substack{k\\k\neq j}}\tan(x_{jk})\Lambda_{jk} .
\eeq    
For the system with periodic boundary condition, Eq.(\ref{hc1}) is true 
for $\mu_1(\lambda)=\lambda,1-\lambda$ and $\mu_2(\lambda)=\lambda,1-\lambda$
whereas for anti-periodic boundary condition, the same form of momentum
operators demand  $\mu_1(\lambda)=\lambda,1-\lambda$ and 
$\mu_2(\lambda)=\frac{1}{2}[1\pm\sqrt{1+4\lambda-4\lambda^2}]$.
From the above 
values of  $\mu_1(\lambda)$ and $\mu_2(\lambda)$ it is observed that for
bosonic ($\lambda=0$) and fermionic ($\lambda=1$) limits, the spectra of
periodic and anti-periodic models are identical. There does not exist any 
other $\lambda$, for which such an identical spectrum is obtained for
periodic and anti-periodic models.
This can be readily checked by putting $\lambda=0,1$ in the respective 
Hamiltonians.

It is further noted that the spectrum for anti-periodic case is real
for a restricted range of values of $\lambda$, i.e.,
$\frac{1}{2}-\frac{1}{\sqrt{2}}\leq \lambda\leq \frac{1}{2}+\frac{1}{\sqrt{2}}$
unlike its periodic counterpart. 

\section{The Calogero-Sutherland model with hyperbolic type interaction}
The hyperbolic form of the CSM may be obtained by taking the limit of
the Weierstrass $\wp$-function which is a 
doubly periodic even elliptic function. It is an analytic function except 
at points which are double poles congruent to the vertices of the period 
parallelogram.
Let us consider the Weierstrass $\sigma$-function given as\cite{cop}
\beq\label{sigma}
\sigma (x) = \frac{2 P_1}{\pi} \exp \left ( \frac{\eta_1 x^2}{
2 P_1} \right )
\sin \left ( \frac{ \pi x}{2 P_1} \right )
\prod_{n=1}^{\infty} \left ( \frac{ 1-2 q^{2n} \cos \left( 
\frac{\pi x}{ P_1} \right ) + q^{4n}}{(1-q^{2n})^2 } \right ),
\eeq
where, $P_1$, $P_2$ are the half-period magnitudes, $q = \exp(i\pi P_2/P_1)$ and
$\eta_1 = \frac{d}{dz}\ln\sigma(z) \vert_{P_1}$.   
The $\wp$-function with period $2 P_1 $ and $2 P_2 $ is given by
\beq\label{pfunc}
\wp (x | 2P_1, 2P_2) = - \frac{d^2}{dx^2}\ln \sigma (x).
\eeq

\subsection{Hyperbolic extension of the CSM with periodic and anti-periodic boundary conditions}
The CSM Hamiltonian with elliptic type interaction under periodic
and anti-periodic boundary condition is written as 
\beq\label{ellip}
H_{N}^e = \sum_{j=1}^{N} {\partial_j}^2
-\lambda (\lambda -1) {\sum_{j,k}}^\prime 
[\wp (x_{jk} | 2P_1, 2P_2)
\pm \wp (x_{jk} + P_1 | 2P_1, 2P_2)] .
\eeq 
The plus and minus sign represent the periodic and anti-periodic cases
respectively. 
In view of Eq.(\ref{pfunc}) and Eq.(\ref{sigma}) 
\beqar\label{lm}
&&\wp (x | 2P_1, 2P_2) |_{q \rightarrow 0} = 
\frac{\pi^2}{4P_1^2}\csc ^2 \left ( \frac{ \pi x}{2 P_1} \right ) + C_1,
\hskip .2cm \mbox{and} \nonumber \\
&&\wp (x +P_1| 2P_1, 2P_2) |_{q \rightarrow 0} = 
\frac{\pi^2}{4P_1^2}\sec ^2 \left ( \frac{ \pi x}{2 P_1} \right ) + C_2,
\eeqar
where $C_1$ $C_2$ are constants, which will henceforth
be ignored as they signify a
trivial constant shift in the spectrum of the Hamiltonian.
Putting $P_1=L$ in Eq.(\ref{lm}) and using the result in Eq.(\ref{ellip}) we get
the trigonometric Hamiltonian. 
On the other hand, replacing $P_1$ by $iL$ we get
\beqar\label{imag}
&&\wp (x | 2iL, 2P_2) |_{q \rightarrow 0} = 
\frac{\pi^2}{4L^2}\csc \! \mbox{h} ^2 \left ( \frac{ \pi x}{2L} \right ),
\hskip .2cm \mbox{and} \nonumber \\
&&\wp (x +iL| 2iL, 2P_2) |_{q \rightarrow 0} =
-\frac{\pi^2}{4L^2}\sec \!\mbox{h} ^2 \left ( \frac{ \pi x}{2L} \right ) .
\eeqar
Changing the variable $(\pi/{2L})x_j\rightarrow x_j $
we get the hyperbolic extension of the periodic Hamiltonian 
as
\beq\label{hyp}
H_N^{h^-}=\sum _{j=1}^{N}{\partial_j}^2-\lambda
(\lambda-1)\left({\sum_{j,k}}^\prime\frac{1}{\sinh^2(x_{jk})}-
{\sum_{j,k}}^\prime\frac{1}{\cosh^2(x_{jk})}\right) .
\eeq
Following calculation similar to that in the 
trigonometric anti-periodic case (Sec.\ref{tacsm}), 
the hyperbolic Hamiltonian in the anti-periodic model can be 
written in the following form,
\beq\label{hyp2}
H^{h^+}_N=\sum _{j=1}^{N}{\partial_j}^2-\lambda
(\lambda-1)\left({\sum_{j,k}}^\prime\frac{1}{\sinh^2(x_{jk})}+
{\sum_{j,k}}^\prime\frac{1}{\cosh^2(x_{jk})}\right) .
\eeq
The above two Hamiltonians can be written in the following compact form
\beq\label{htrig_gen2}
H_N^{h^{\mp}}=\sum _{j=1}^{N}{\partial_j}^2-\lambda
(\lambda-1){\sum_{j,k}}^\prime[V(x_{jk})\mp R(x_{jk})] ,
\eeq
where the minus sign corresponds to the periodic case and the plus 
sign to the anti-periodic case.

\subsection{Gauge momentum operators for the hyperbolic extensions}
Using $v(x)$ and $r(x)$ as defined in Eq.(\ref{vr}), we can construct 
the momentum operators for the hyperbolic models. One may easily
verify that $r(x)\cdot v(x)=1$. The gauge momentum operators must satisfy
the following properties to ensure the integrability of the
hyperbolic models.
\beq\label{c0}
[\Lambda_{ij},d_k]=0  .
\eeq
\beq\label{c1}
\sum_{j=1}^N d_j^2= H^{h^{\mp}}_N \, + \, \mbox{constant}  .
\eeq
\beq\label{c2}
[d_j,d_k]=0 .
\eeq
\beq\label{c3}
[d_j,H_N^{h^{\mp}}]=0 .
\eeq
The above properties in turn imply that $v(x)$ and $r(x)$ be such that
\beqa
\frac{d}{dx}v(x)=V(x) \; ,& \hskip 2cm v^2(x)=V(x)+\mbox{constant} \\
\frac{d}{dx}r(x)=-R(x) \; ,& \hskip 2cm  r^2(x)=-R(x)+\mbox{constant}
\eeqa
From the above, we may write the momentum operators for the hyperbolic
models as 
\beq\label{mom2}
d_j={\partial _j}-\mu_1(\lambda)\sum_{\substack{k \\ k\neq j}}
\coth(x_{jk})
\Lambda_{jk}-\mu_2(\lambda)\sum_{\substack{k \\ k\neq j}}
\tanh(x_{jk})\Lambda_{jk} .
\eeq    
The dependence of $\mu_1(\lambda)$ and $\mu_2(\lambda)$ on $\lambda$ 
and the restrictions imposed on the values of $\lambda$ is
similar to that obtained for trigonometric models.

\section{Conclusion}
In this article, we have studied the classical Calogero system 
with anti-periodic boundary condition. This system is 
equivalent to a Calogero system with periodic boundary condition in the
presence of a transverse magnetic field. 
Though the presence of a magnetic field makes the system physically
interesting, the anti-periodic Calogero systems are not widely studied.
The reason perhaps
lies in the fact that certain algebraic symmetries based on the root systems
present in the periodic case are
not available in the anti-periodic case. This makes the anti-periodic
CSM more difficult and involved. 

In this article, the integrability of the Calogero system 
with anti-periodic boundary
condition has been established by constructing a family of 
commuting momentum operators. We have shown that,
for certain restrictions on the
values of the interaction parameter $\lambda$, the momentum 
operators are hermitian.
Relevant extensions of the
system, both classical and spin cases, should be an interesting subject
of future study for investigating integrability and
solvability.

\section{appendix}

The commutation property of the gauge momentum operators is explicitly
demonstrated in this appendix.

The general form of a gauge momentum operator is,
\beq\label{typmom}
d_j = \partial_j + \mu_1 \sum_{\substack{m\\ m\neq j}} X_{jm}
\Lambda_{jm} + \mu_2 \sum_{\substack{n\\ n\neq j}} Y_{jn}
\Lambda_{jn} .
\eeq
Here, $X$ and $Y$ are odd trigonometric or hyperbolic functions and hence
their derivatives, $X'$ and $Y'$ are even functions. 

Let us consider the successive action of $d_k$ and $d_j$ 
on $\psi$ (where $j \neq k$),
\beqar\label{apmom1}
d_j d_k \psi && = [\partial_j + \mu_1 \sum_{\substack{m\\ m\neq j}}X_{jm}
\Lambda_{jm} + \mu_2 \sum_{\substack{n\\ n\neq j}} Y_{jn}
\Lambda_{jn}] [ \partial_k \psi  + \mu_1 \sum_{\substack{r\\ r\neq k}} X_{kr}
\psi + \mu_2 \sum_{\substack{s\\ s\neq k}} Y_{ks}\psi] \nonumber \\
&& = \partial_j \partial_k \psi + \mu_1 \sum_{\substack{r\\ r\neq k}} X_{kr}
\partial_j \psi + \mu_1 X'_{kj}\psi + \mu_2 \sum_{\substack{s\\ s\neq k}} 
Y_{ks} \partial_j \psi + \mu_2 Y'_{kj} \psi +\mu_1 X_{jk} \partial_j 
\psi \nonumber \\
&& + \mu_1 \sum_{\substack{m \\ m\neq \{j , k\}}}
X_{jm} \partial_k \psi + \mu_1^2 \sum_{\substack{m\\ m\neq j}}X_{jm}
\Lambda_{jm}\sum_{\substack{r\\ r\neq k}} X_{kr}\psi + \mu_1  \mu_2
\sum_{\substack{n\\ n\neq j}} X_{jm} \Lambda_{jm} \sum_{\substack{s\\ s\neq k}}
Y_{ks}\psi + \mu_2 Y_{jk} \partial_j \psi \nonumber  \\
&& + \mu_2 \sum_{\substack{n\\ n\neq \{j, k\}}} Y_{jn} \partial_k \psi + 
\mu_1 \mu_2 \sum_{\substack{n\\ n\neq j}} Y_{jn} \Lambda_{jn}
\sum_{\substack{r\\ r\neq k}} X_{kr}\psi + \mu_2^2 
\sum_{\substack{n\\ n\neq j}} Y_{jn}\Lambda_{jn}
\sum_{\substack{s\\ s\neq k}} Y_{ks}\psi.
\eeqar
Now consider the successive action of $d_j$ and $d_k$ on $\psi$
\beqar\label{apmom2}
d_k d_j \psi && = \partial_k \partial_j\psi + \mu_1 
\sum_{\substack{m\\ m\neq j}}
X_{jm}\partial_k \psi + \mu_1 X'_{jk} \psi + \mu_2
\sum_{\substack{n\\ n\neq j \neq k}} Y_{jn} \partial_k \psi + \mu_2 Y'_{jk}
\psi \nonumber \\
&& +\mu_1 X_{kj} \partial_k\psi +  
\mu_1\sum_{\substack{r\\ r \neq \{k, j\}}} X_{kr}
\partial_j\psi + \mu_1^2 \sum_{\substack{r\\ r\neq k }}X_{kr}\Lambda_{kr}
\sum_{\substack{m\\ m\neq j}} X_{jm} \psi + 
\mu_1\mu_2 \sum_{\substack{r\\ r\neq k}}X_{kr}\Lambda_{kr} 
\sum_{\substack{n\\ n\neq j}} Y_{jn}\psi \nonumber \\
&& + \mu_2 Y_{kj}
\partial_k\psi + \mu_2 \sum_{\substack{s\\ s \neq \{k, j\}}} Y_{ks} 
\partial_j\psi
+ \mu_1\mu_2 \sum_{\substack{s\\ s\neq k }} Y_{ks}\Lambda_{ks}
\sum_{\substack{m\\ m\neq j}} X_{jm}\psi \nonumber \\
&& + \mu_2^2  
\sum_{\substack{s\\ s\neq k }} 
Y_{ks}\Lambda_{ks}\sum_{\substack{n\\ n\neq j}} Y_{jn}\psi .
\eeqar
In the commutator the first, third and fifth terms 
of Eq.(\ref{apmom1}) cancel the corresponding terms of Eq.(\ref{apmom2}). 
Of the remaining terms, collecting the terms containing 
the first order derivative of $\psi$ in Eq.(\ref{apmom1}) and 
Eq.(\ref{apmom2}), we may write
\beqar\label{a}
&& \mu_1 X_{kj}\partial_j\psi + \mu_1\sum_{\substack{r\\ r\neq \{k, j\}}}
X_{kr}\partial_j\psi + \mu_2 Y_{kj}\partial_j\psi + \mu_2 
\sum_{\substack{s\\ s\neq \{k, j\}}} Y_{ks}\partial_j\psi \nonumber \\ 
&& + \mu_1 X_{jk}
\partial_j\psi + \mu_1\sum_{\substack{m\\ m\neq \{k, j\}}} X_{jm}\partial_k
\psi +\mu_2 Y_{jk} \partial_j\psi + \mu_2\sum_{\substack{n\\ n\neq \{k,j\}}}
Y_{jn}\partial_k\psi
\eeqar
\beqar\label{b}
&& \mu_1 X_{jk}\partial_k\psi + \mu_2\sum_{\substack{m\\ m\neq \{k, j\}}}
X_{jm}\partial_k\psi + \mu_2 Y_{jk}\psi + 
\mu_2\sum_{\substack{n\\ n\neq \{k, j\}}}
Y_{jn}\partial_k\psi  \nonumber \\
&& +\mu_1 X_{kj} \partial_k\psi + \mu_1 
\sum_{\substack{r\\ r\neq \{k,j\}}} X_{kr} \partial_j\psi + \mu_2 Y_{kj}
\partial_k\psi + \mu_2\sum_{\substack{s\\ s\neq \{k, j\}}} Y_{ks}
\partial_j\psi
\eeqar
In the above two expressions the first and the fifth terms cancel 
each other as $X$ is an odd function. The same holds true for 
the third and seventh terms. 

Finally the remaining terms in expression (\ref{a}) i.e.,
the second, fourth, sixth and eighth cancel with
the sixth, second, eighth and fourth terms respectively 
of expression (\ref{b})
upon commutation.

Now we are left with the terms containing 
$\mu_1^2, \mu_2^2$ and $ \mu_1 \mu_2$
in the expression of the commutator. The coefficient of $\mu_1^2$ is
given by 
\bd
\sum_{\substack{m\\ m\neq j}} X_{jm} \Lambda_{jm} 
\sum_{\substack{r\\ r\neq k}}
X_{kr} \psi - \sum_{\substack{r\\ r\neq k}} X_{kr} 
\Lambda_{kr} \sum_{\substack{m\\ m\neq j}} X_{jm} \psi .
\ed 
This term vanishes because of the symmetry in the indices $j$ and $k$. By the
same argument the coefficients of $\mu_2^2$ and $\mu_1\mu_2$ also
vanish.
Hence it is proved that the gauge momentum operators $d_j$, $d_k$ mutually
commute for $j,k = 1, \dots N$.

\section*{Acknowledgment}

AC wishes to acknowledge the Council of Scientific and
Industrial Research, India (CSIR) for fellowship support.

%

\end{document}